\documentclass[preprint,showpacs,prl]{revtex4}

\usepackage{graphicx,amssymb,amsmath}

\usepackage{times}

\begin{document}

\title{Diffusion and spatial correlations in suspensions of swimming particles}

\author{Patrick T. Underhill}%

\author{Juan P. Hern\'andez-Ortiz}%
\altaffiliation[Current Address: ]{Departamento de Materiales,
Faculdad de Minas , Universidad Nacional de Colombia, Medellin ,
Carrera 80, \#65-223, Bloque M3-050 Medellin, Colombia}

\author{Michael D. Graham}%
 \email[Corresponding author: ]{graham@engr.wisc.edu}
\affiliation{%
Department of Chemical and Biological Engineering, University of
Wisconsin-Madison,\\ 1415 Engineering Dr., Madison, Wisconsin 53706,
USA}%

\date{\today}

\begin{abstract}
Populations of swimming microorganisms produce fluid motions that
lead to dramatically enhanced diffusion of tracer particles. Using
simulations of suspensions of swimming particles in a periodic
domain, we capture this effect and show that it depends
qualitatively on the mode of swimming: swimmers ``pushed'' from
behind by their flagella show greater enhancement than swimmers that
are ``pulled'' from the front. The difference is manifested by an
increase, that only occurs for pushers, of the diffusivity of
passive tracers and the velocity correlation length with the size of
the periodic domain.  A physical argument supported by a mean field
theory sheds light on the origin of these effects.
\end{abstract}


\maketitle

In the present work we consider the possibility of emergence of
large scale fluid motion and enhanced fluid transport in populations
of small swimming organisms. At the global scale, it has been
suggested that swimming organisms such as krill can alter mixing in
the oceans~\cite{KunzeKrillBioTurbulence06,VisserNoBiomixing07}. At
the laboratory scale, experiments with suspensions of swimming cells
have revealed characteristic swirls and jets much larger than a
single cell, as well as causing increased diffusivity of tracer
particles~\cite{MendelsonWhirls99,WuLibchaber00,GoldsteinKesslerSPP04,RiedelSperm05,AransonSwimExpts07}.
This enhanced diffusivity may have important consequences for how
cells reach nutrients, as it indicates that the very act of swimming
towards nutrients alters their distribution. The enhanced
diffusivity has also been proposed as a scheme to improve transport
in microfluidic devices~\cite{BreuerSPPdiffusion04} and might be
exploited in microfluidic cell culture of motile organisms or cells.

The feedback between the motion of swimming particles and the fluid
flow generated by that motion is thus very important. Nevertheless,
in the literature on collective dynamics of self-propelled
particles, this effect has received little attention. Most previous
attempts to understand the collective motion of swimming
micro-organisms fall into two broad categories: ``particle level''
approaches based on \emph{ad hoc} interaction rules between the
individual agents, e.g.~\cite{VicsekSPP95}, and continuum models of
active suspensions based on phenomenological field
equations~\cite{RamaswamySPP02,KruseAstersPRL04,AransonFieldModel07}.
The hydrodynamics of single swimming organisms at low Reynolds
number have been studied for many
years~\cite{TaylorSwimmers51,Lighthill76,PurcellAJP77}, and studies
of hydrodynamic interactions (HI) have also been performed for pairs
of
swimmers~\cite{PhanThienComputMech97,PedleyPairInteract2006,IshikawaBioPhysJ07}.
However, only a small number of
studies~\cite{HernandezPRL05,StoltzThesis06,PagonabarragaEPL06,ShelleyRodsPRL07,PedleyRheologyJFM2007,PedleyDiffusionJFM2007}
have examined particle level models of \emph{populations} of
swimmers that include the fluid motion caused by the swimmers and
its influence on transport. Hern\'andez-Ortiz et
al.~\cite{HernandezPRL05,StoltzThesis06} developed a simple physical
model of self-propelled particles, and performed simulations of them
in a confined domain. Their work showed that multi-body HI between
self-propelled particles were sufficient to lead to fluid motions
characteristic of those seen in experiments. Llopis and
Pagonabarraga~\cite{PagonabarragaEPL06} performed simulations of
self-propelled particles in two-dimensions, observing aggregates of
swimming particles and a decrease of velocity with increasing volume
fraction. Saintillan and Shelley~\cite{ShelleyRodsPRL07} performed
simulations of long, slender self-propelled rods propelled by an
imposed shear stress on the surface of part of the rods. Their main
result was that initially nematically ordered swimmers do not remain
aligned except with very near neighbors. Pedley and
coworkers~\cite{PedleyRheologyJFM2007,PedleyDiffusionJFM2007}
examined the rheology and random walk motion of semi-dilute
suspensions of self-propelled spheres.

We consider here a suspension of $N$ neutrally buoyant rodlike
swimmers in a spatially periodic cubic fluid domain of side length
$L$, in a concentration regime dilute enough that the dominant
interactions between the particles arise from the fluid motions that
they generate as they swim. Each swimmer has a characteristic length
$\ell$, and in isolation moves in a straight line with speed
$v_{is}$. It is assumed that the Reynolds number, $v_{is} L / \nu $,
is much less than 1, where $\nu$ is the fluid kinematic viscosity.
The concentration is given as an effective volume fraction $\phi_e =
\pi N \ell^3 / (6 L^3)$ -- this would be the true volume fraction if
the swimmers were spheres of diameter $\ell$. To allow large
populations ($> 10^4$ swimmers) over long times, a simple model of
each swimmer is adopted, following our previous
work~\cite{HernandezPRL05}.

\begin{figure}
\begin{center}
\includegraphics[width=3.25in]{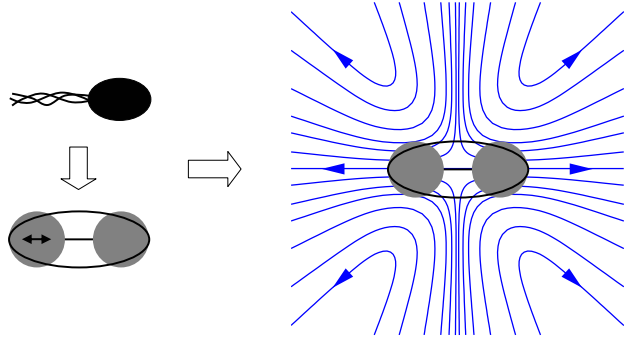}
\caption{(color online) Illustration of a pushing organism, swimmer
model, and fluid disturbance they cause. The double-arrow signifies
the flagellum force acting on the bead and opposite force acting on
the fluid, both acting at the center of the first bead. The blue
lines represent streamlines of the axisymmetric fluid disturbance. A
puller produces the same streamlines with the arrows reversed. }
\end{center}
\end{figure}

Each self-propelled particle is represented as two beads connected
by a stiff spring with equilibrium length $\ell$ as shown in Figure
1. The unit vector pointing along the axis from bead 1 to bead 2 is
denoted $\mathbf{n}$. The propulsion is provided by a ``phantom
flagellum'' that we do not treat explicitly, but only through its
effect on the swimmer body and the fluid. This flagellum exerts a
force $\mathbf{F}_{f}$ on bead 1 of the swimmer, and also exerts a
force $-\mathbf{F}_{f}$ on the fluid. This force on the fluid occurs
at the position of bead 1. With this model we can consider
``pushers'' or ``pullers'' depending on whether $\mathbf{F}_{f}$ is
parallel or anti-parallel to $\mathbf{n}$, respectively. A pusher
sends fluid away from it fore and aft, with fluid moving toward its
``waist'', and vice versa for a puller. A cell whose flagella propel
it forward predominantly from behind would be a
pusher~\cite{BrayBook01,HernandezPRL05}. We shall see that this
distinction is important. The swimmers also interact with an
excluded volume potential using the repulsive portion of the
Gay-Berne potential~\cite{AllenGermanoMolPhys06}.

The motion of the swimmers is calculated by writing a force balance
(neglecting inertia because of the small size of a microorganism)
for each bead. On bead 1 of each swimmer this is $ \mathbf{F}_{f} +
\mathbf{F}_{h1} + \mathbf{F}_{c1} + \mathbf{F}_{e1} = \mathbf{0} , $
where $\mathbf{F}_{h1}$ is the hydrodynamic drag force on the bead,
$\mathbf{F}_{c1}$ is the connector (spring) force on the bead, and
$\mathbf{F}_{e1}$ is the excluded volume force on the bead. On bead
2 of each swimmer the force balance is identical except without a
flagellum force. The drag force is written using Stokes law,
$\mathbf{F}_{h1} = -6 \pi \eta a [ \dot{\mathbf{r}}_1 -
\mathbf{v}'(\mathbf{r}_1) ] $, where $a$ is the bead radius, $\eta$
is the fluid viscosity , $\mathbf{r}_1$ is the position of the bead,
and $\mathbf{v}'(\mathbf{r}_1)$ is the fluid velocity at
$\mathbf{r}_1$ generated by all other beads and phantom flagella.
This fluid velocity is calculated using an order $N$
method~\cite{HernandezPRL07}, treating each bead of the swimmer as a
regularized point force. We set $\ell = 3 a$. Note that the net
force exerted by an isolated swimmer on the fluid is zero --
overall, a neutrally buoyant swimmer is a force dipole to leading
order. This model captures the universal far-field behavior while
neglecting the near-field corrections to interactions between
swimmers, which are dependent on the details of the organism. The
validity of this approximation is supported by recent
simulations~\cite{NottJFM2008}. Finally, we note that the limited
set of results with multi-bead rod swimmers is qualitatively
consistent with those for the two-bead swimmers.

The fluid motion generated by each swimmer perturbs the trajectories
of fluid elements (tracers) and other swimmers. At the volume
fractions considered here, this motion also eliminates any long
range orientational order among the swimmers. The motion of both
tracers and swimmers, while ballistic at short times, becomes
diffusive at long times. The tracer diffusion in particular shows
important and unanticipated behavior. Figure 2 shows tracer mean
squared displacements (inset) and the long-time tracer diffusivity
$D_t$ in a suspension of pushers as a function of system size for
$\phi_{e}=0.1$. The diffusivity grows significantly with system
size, greatly enhancing it beyond that for a suspension of pullers
(also shown), which exhibits only weak or negligible system size
dependence.

The system size dependence observed for pushers is due to collective
behavior; simulations in which the swimmers do not interact,
swimming in straight paths, do not show a significantly changing
diffusivity. This observation has important implications for
transport of nutrients, chemoattractants, and other chemical species
in the environment of a population of swimming organisms. For
example, organisms which concentrate toward an attractant will
disperse that attractant more rapidly if they are pushers than if
they are pullers. The remainder of this report describes efforts to
gain an understanding of the increased transport.

The diffusivity is related to the fluid velocity through the
Green-Kubo relation~\cite{McQuarrieBook},$ D_{t} = ( 1/3 )
\int_0^{\infty} \langle \mathbf{v}_{t}(0) \cdot \mathbf{v}_{t}(t)
\rangle dt , $ where $\mathbf{v}_{t}$ is the velocity of a tracer
(fluid element) and angle brackets indicate an ensemble average. The
mean squared tracer velocity $\langle v_{t}^{2}\rangle=\langle
\mathbf{v}_{t}(0) \cdot \mathbf{v}_{t}(0)\rangle$ is shown in Fig. 2
and has a much weaker system size dependence. The simple idea that
the swimmer velocity is the isolated value plus the tracer velocity,
$ \mathbf{v}_s = \mathbf{v}_{is} + \mathbf{v}_t$ , predicts that
$\langle v_{s}^{2}\rangle \approx v_{is}^{2}+\langle
v_{t}^{2}\rangle $, which is verified in Fig. 2.

\begin{figure}
\begin{center}
\includegraphics{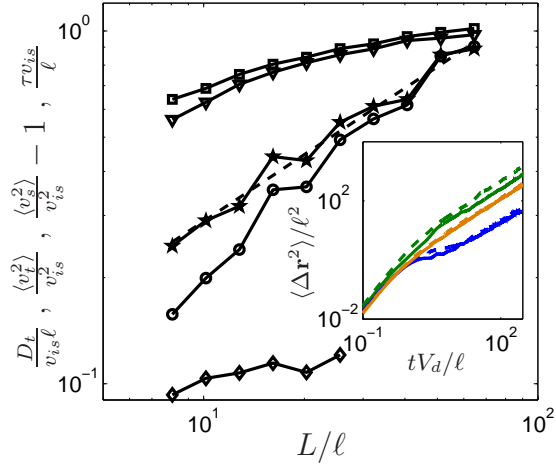}
\caption{(color online) System size dependence at volume fraction
$\phi_{e}=0.1$ for pushers. Variation of mean squared tracer
velocity $\langle v^{2}_t \rangle$ (squares), $\langle
v_{s}^{2}\rangle-v_{is}^2$ (triangles), tracer diffusivity $D_{t}$
(circles), and tracer correlation time (stars). The dashed line is a
power law with $\alpha=0.63$. The tracer diffusivity in a suspension
of pullers is shown with diamonds. (inset) Tracer mean-squared
displacements versus time for independent swimmers (blue), pullers
(orange), and pushers (green) for $N=400$ (solid) and $N=3200$
(dashed). Averages over $400$ tracers typically give statistically
converged results.}
\end{center}
\end{figure}

We now define a correlation time, $\tau = D_{t} / \langle v_{t}^2
\rangle$, for the trajectories of tracer particles -- tracer motion
becomes diffusive on times $t\gg \tau$. Figure 2 shows $\tau$ versus
$L$; over the entire range of system sizes considered, this relation
displays a clear power law dependence $\tau\sim L^{\alpha}$, with
$\alpha= 0.63 \pm 0.03$, over a range spanning a factor of  $8$ in
$L$ and thus $512$ in $N$. This data represents the first
examination of system size effects on the fluid response in swimming
suspensions, and includes orders of magnitude more swimmers than any
previous studies. For larger systems the behavior may transition to
a non power law regime, but the system size dependence shown here
will nevertheless cause significant enhancement of transport in a
suspension of pushers.

The time $\tau$ represents the time a tracer moves in the same
direction. A corresponding step length or correlation length $l_{t}$
for the tracer's random walk scales as $ \tau \langle
v_{t}^{2}\rangle^{1/2}$. Because $\langle v_{t}^{2}\rangle$ has a
much weaker dependence on $L$ than $\tau$, our results imply the
existence of a correlation length for pushers that obeys $l_{t}\sim
\ell (L/\ell)^{\alpha} \sim \ell^{(1-\alpha)} L^{\alpha}$ over the
range of system sizes examined. The suspension of pushers leads to
enhanced correlation lengths in the fluid, beyond pullers or
independent swimmers, because of this increase with system size.

Before elaborating on correlation lengths, we briefly address the
concentration dependence, using simple scaling arguments. In the
Green-Kubo relation, the tracer velocity is a sum of disturbances
due to each swimmer. Assuming independence of the swimmers in the
dilute limit, the cross terms from the dot product vanish, resulting
in $D_t \sim \phi_e$. For the same reason, $\langle v_{t}^{2}\rangle
\sim \phi_e$, which leads to a correlation time $\tau$ that is
independent of volume fraction. Turning to swimmers rather than
tracers, we note that in the dilute limit, the swimmers continue in
straight lines until perturbed by other swimmers. If we define a
cross section $\sigma$ for these ``collisions'', the mean free path
or step length $l_s$ scales as $\ell^3/(\sigma \phi_e)$. Because
$\langle v_{s}^{2}\rangle \approx v_{is}^{2}+\langle
v_{t}^{2}\rangle $ and $ \langle v^2_t \rangle \sim \phi_e  $, the
root-mean-squared swimmer velocity is $\langle v^2_s \rangle^{1/2}
\sim v_{is} + O(\phi_e)$. Therefore, the diffusivity scales as $D_s
\sim \langle v^2_{s} \rangle^{1/2} l_s \sim \phi_e^{-1}$. For
$\phi_e \lesssim 10^{-2}$ simulation results follow these
predictions (not shown), while above that volume fraction, they
begin to deviate, indicating a breakdown of the assumption of
independent swimmers.

We now return to the issue of correlation lengths, and consider the
fluid velocity spatial correlation function, $ C(\mathbf{r}) = \left
\langle L^{-3} \int \mathbf{v}_t(\mathbf{r}') \cdot
\mathbf{v}_t(\mathbf{r}' + \mathbf{r}) d \mathbf{r}'  \right \rangle
$. This is shown in Figure 3 at a volume fraction of $\phi_e = 0.1$
for a range of system sizes. For independently distributed point
dipoles, this correlation is $ C_{id}(\mathbf{r}) = V^2_{d} \ell/r
$, where $V^2_{d} = \phi_e d^2 / ( 5 \eta^2 \pi^2 \ell^4 ) $, and
$d$ is the dipole moment~\cite{VsqTrIndependTheoryNote}. Because of
the finite size of the swimmers and domain, real independent
swimmers will deviate from this relation when $r \lesssim \ell$ or
$r\approx L$. Even this simple result illustrates the possible
complexities that arise in considering swimming suspensions; the
velocity correlations are very long-ranged even in the absence of
correlations between the positions and orientations of the swimmers.

\begin{figure}
\begin{center}
\includegraphics{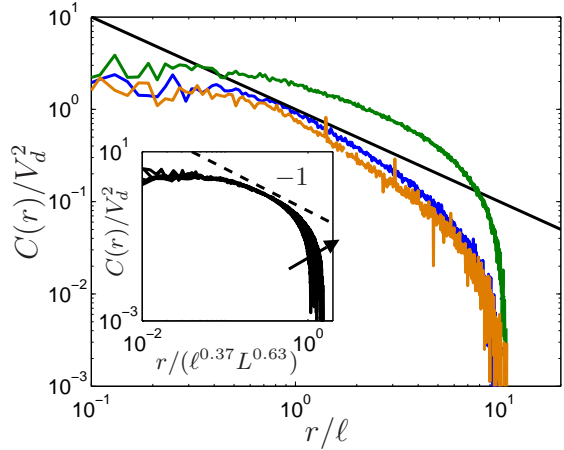}
\caption{(color online) The correlation function at $N=3200$ for
different models at $\phi_{e}=0.1$: independent swimmers (blue),
independent point dipole theory (black), full simulation of pushers
(green), and pullers (orange). (inset) Correlation function for
pushers at different system sizes with the spatial extent rescaled
by $l_t = \ell^{0.37} L^{0.63}$. The arrow denotes increasing system
sizes $N=400 \to 6400$.}
\end{center}
\end{figure}

Figure 3 shows the correlation function at a characteristic system
size of $N=3200$ ($L=25.589 \ell$), for pushers, pullers,
noninteracting pushers (where the swimmers do not feel the fluid
motion but the tracer particles do) and the point dipole theory
result. The noninteracting swimmer simulation shows the expected
deviations from $C_{id}$ at small $r$ and large $r$ due to the
assumptions of the theory. For the interacting pushers, the tracer
correlation is larger than that of the ideal theory and is system
size-dependent as illustrated below, while for pullers, it remains
close to that for independent swimmers, regardless of system size.
The presence or absence of system size dependence in the velocity
correlations is consistent with that for tracer diffusivity.

The connection between extended spatial correlations and the system
size dependent diffusivity for pushers is made clearer by returning
to the system size dependent correlation length $l_{t}$. The inset
to Fig. 3 shows the velocity correlation function for pushers at
$\phi_{e}=0.1$ for several different system sizes with distance
scaled by $l_{t}$. Except at distances close to the total box size,
this scaling collapses the results to a single curve. This suggests
that the correlation length inferred from the diffusivity also
controls the spatial velocity correlations. At intermediate
distances, the curves are consistent with a generalized point dipole
theory, with a correlation of the form $C(r) \sim
L^{\alpha}\ell^{(1-\alpha)}/ r $ .

The origin of the difference between pusher and puller dynamics is
not completely clear, but a simple argument and some analysis can
shed some light on the issue.  Imagine an initially homogeneous
isotropic suspension of swimmers subject to a perturbation in the
form of a shear flow. In a suspension of pullers, the shear flow
distorts the orientation distribution and the resulting shear stress
due to this distortion opposes the original shear flow, driving the
system back to isotropy.  This situation is closely analogous to a
Brownian suspension of fibers or polymer molecules, the difference
being that in the case of the swimmers the force dipoles that lead
to the stress are intrinsic rather than induced by the shear. In the
case of pushers, however, the initial shear flow distorts the
orientation distribution in the same way, but now the force dipoles
have opposite signs from the puller case and lead to a shear stress
that \emph{enhances} the original shear flow perturbation. This
enhancement further increases the orientation of the swimmers thus
implying instability with respect to shear perturbations of the
homogeneous isotropic state.

This argument can be made precise through a simple mean field theory
(a related analysis without the above physical argument is given
by~\cite{ShelleyKineticTheoryPRL08}). Let
$\mathbf{u}=\langle\mathbf{v}_{t}\rangle$ and
$\boldsymbol{\alpha}=\langle \mathbf{nn}\rangle$, where $\mathbf{n}$
is the director vector of a swimmer.  For a homogeneous isotropic
state, the number density $c$ is constant,
$\mathbf{u}=\langle\mathbf{n}\rangle=0$ and
$\boldsymbol{\alpha}=\frac{1}{3}\boldsymbol{\delta}$. In the point
dipole limit the stress tensor generated by the swimmers is
$\boldsymbol{\tau}_{s}=dc\boldsymbol{\alpha}$, where $d$ is the
dipole strength per swimmer ($d>0$ for pullers, $d<0$ for pushers).
Stokes' equation relates $\mathbf{u}$ and $\boldsymbol{\tau}_{s}$,
and $\mathbf{n}$ evolves as would an infinitesimal material line
subject to the constraint of constant length~\cite{DPL2}.
Considering linear stability of the homogeneous isotropic state
subject to a shear flow disturbance $u_{y}(x)$ in the long wave
limit, Stokes' equation and the evolution equation for $\alpha_{xy}$
become $ \eta \partial_{x} u_{y} = -d c \alpha_{xy} $ and $
\partial_{t} \alpha_{xy} = (1/5) \partial_x u_{y} $.
Combining these shows that $\alpha_{xy}$ evolves exponentially in
time with growth rate $\sigma={-dc}/{5\eta}$; for  pullers, the
shear perturbation decays, while for pushers it grows, confirming
the simple physical argument given above. This result is
wavelength-independent in the long-wave limit and illustrates a
mechanism for generating long-range correlations as seen in the full
pusher simulations.

The qualitative differences we observe between pushers and pullers
naturally leads to the question of whether different organisms have
evolved their method of swimming partially based on how
significantly that method of swimming enhances transport in the
fluid. In addition to biological systems, a number of artificial
micro-swimmers have recently been
developed~\cite{WhitesidesACIE2002,GolestanianPRL05,StoneMagSwimmerNature2005,PaxtonLocoReview2006,HoggRobot07}.
The impact shown here of the mode of swimming on collective behavior
could be an important design criterion for future devices of this
kind.

We gratefully acknowledge support from NSF grants CTS-0522386 and
DMR-0425880.


\begin{thebibliography}{37}
\bibitem[{Vsq()}]{VsqTrIndependTheoryNote}
\bibinfo{note}{Our independent dipole theory gives a mean-squared tracer
  velocity $\langle v^2_t \rangle = C (0) \sim (2 \phi_{e} d^2 k_{cut} )/(5
  \eta^2 \pi^3 \ell^3)$, where $k_{cut} \approx 2 \pi / \ell$ represents the
  characteristic wavevector beyond which the point dipole theory is
  regularized.}
\end{thebibliography}

\end{document}